\documentclass{article}
\usepackage{ltwol2e}

\def\.{\cdot}

\def\be{\begin{eqnarray}}
\def\nn{\nonumber\\}
\def\ee{\end{eqnarray}}
\def\({\left(}
\def\[{\left[}
\def\){\right)}
\def\]{\right]}
\def\h{{1\over 2}}

\begin{document}

\title{QCD PHASE SHIFTS AND RISING TOTAL CROSS SECTIONS}

\author{C.S. Lam}

\address{Department of Physics, McGill University, 3600 University
St., Montreal, QC, Canada H3A 2T8\\E-mail: Lam@physics.mcgill.ca}   

\twocolumn[\maketitle\abstracts{Energy dependence of $\gamma^*p$
total cross section is considered. It is known from the HERA data
that the cross section grows with energy, and the rate of growth is
an increasing function of the virtuality $Q^2$ of the photon $\gamma^*$.
This dependence can be explained in a simple and generic way
by the Froissart bound. To implement this mechanism quantitatively
a theory satisfying $s$-channel unitarity is required. This is achieved by
computing the total cross section from the optical theorem, and the
forward elastic amplitude from the phase shifts in the impact-parameter
 representation. A recipe to do so in perturbative
QCD is discussed, together with an expose of the advantage 
to calculate the elastic amplitude via phase shift, rather than
direct sum of all Feynman diagrams.
A two-loop computation of phase shift and total cross section is 
presented, and compared with the HERA data with good agreements.}]

\section{Total Cross Sections}
\subsection{Experimental Data}
$\gamma^*p$ total cross section $\sigma_{tot}(s,Q)$ measured at 
HERA~\cite{ZEUS,H1} can be fitted by the formula 
$\sigma_0(Q)(s/\Lambda^2)^{a(Q)}$, where $s$ is the square of the c.m.
energy, and $Q^2$ is the virtuality of the photon. The power $a(Q)$
is equal to 0.08 at $Q=0$, the same as the corresponding power
for {\it all} hadronic total cross sections~\cite{DL}. 
 We shall refer to this experimental feature, that the energy dependence
of the total cross section is independent of the nature of the
colliding particles, as {\it universality}. 
Moreover, $a(Q)$ is an increasing function of 
$Q$. Asymptotically, 
total cross sections are limited by the Froissart bound not to grow
faster than $\ln^2s$, so the power-like increase observed must be
approximate, with the effective power of growth
decrasing to zero at infinite energies.

\subsection{$Q$-Dependence of $a(Q)$}
This last remark provides a simple and generic mechanism to explain why
$a(Q)$ is an increasing function of $Q$.
From universality, we can write $\sigma_{tot}(s,Q)$
in the form 
\be\sigma_{tot}(s,Q)=\sigma_0(Q)f(s'),\label{f}\ee 
where $\sigma_0(Q)$ depends both on $Q$ and
 the nature of the colliding particles, but $f(s')$, which governs
the energy dependence of the total cross sections, is 
universal and independent of the nature of the colliding particles.
It is a dimensionless
function of the dimensionless energy $s'=s/\Lambda^2(Q)$.
On dimensional grounds, the energy-scale parameter
$\Lambda(Q)$ is expected to increase with $Q$.

To satisfy the Froissart bound, the universal function $f(s')$ must
asymptotically approach $\ln^2s'$.  This means that at sufficiently high
energies, its slope in a
$\ln f(s')$ vs $\ln s'$ plot must increase with decreasing $s'$.
A larger $Q$ leads to a larger $\Lambda(Q)$, a smaller $s'$,
and hence a larger slope $a(Q)$ in the log-log plot. This shows that $a(Q)$
is an increasing function of $Q$, just as observed in the HERA
data~\cite{ZEUS,H1}.

\section{QCD Phase Shifts}
\subsection{Impact-Parameter Representation}
To be able to
implement this mechanism quantitatively, we must start from a theory
capable of describing rising total cross sections
which obey the Froissart bound. This is so if total 
cross sections are computed from the optical theorem, and the forward
elastic scattering ampltudes are calculated from phase shifts in the
impact-parameter representation:
\begin{eqnarray}
\sigma_{tot}(s)&=&{1\over s}{\rm Im} A_0(s,0)\label{opt}\\
A(s,\Delta)&=&-is\int d^2b 
e^{i\vec \Delta\cdot\vec b}\(e^{2i\delta(s,b)}-1\)\label{impact}
\end{eqnarray}
The impact parameter $\vec b$ and its conjugate, the momentum transfer
$\vec\Delta$, are two-dimensional vectors in the transverse
plane. In QCD, the amplitude $A$ is a colour matrix whose matrix
elements are the scattering amplitudes of partons of various colours.
The amplitude $A_0$ appearing in (\ref{opt}) is obtained from $A$
of (\ref{impact}) by retaining only the colour-singlet exchange amplitude.

The effective size $R$ measured in the total cross section 
can be estimated from the condition $\delta(s,R)\sim
1$. If the interaction has a range $\mu^{-1}$, so that $\delta(s,b)\sim\xi(s)\exp(-\mu b)$ for large $b$,
then the effective radius is given by $R=\mu^{-1}\ln\xi(s)$. This grows
with enegy if $\xi(s)$ does, and as long as $\xi(s)$ does not grow faster
than a power, $R$ will not rise more rapidly than $\ln s$, and 
$\sigma_{tot}\sim\pi R^2$ will not grow faster than $\ln^2s$, as
prescribed by the Froissart bound.

\subsection{QCD Phase Shifts}
So far the discussion has been general and qualitative. To proceed further 
we must find a way to calculate the phase shift $\delta(s,b)$. 
We shall discuss how to do so within the framework of perturbative QCD.

Scattering amplitudes are given by Feynman diagrams. From (\ref{impact}),
the phase shift is given by the logarithm of the amplitude in the
impact-parameter space. With the presence of
non-commuting colour matrices in the amplitude this calculation promises
to be complicated. It is not a priori clear that any simple formula
can be obtained for the phase shift. 
Fortunately, in the leading-log approximation (LLA)
where only the highest power of $\ln s$ is kept, the result turns out to
be very simple~\cite{LL1,FHL,FL,JMP,PH1,PH2}. We will now state this result
for quark-quark scattering.

First, use the $SU(N_c)$ commutation relation 
$[\lambda_a,\lambda_b]=if_{abc}\lambda_c$
for colour matrices $\lambda_a$ in any representation, together with the
normalization condition Tr$(\lambda_a\lambda_b)=\delta_{ab}/2$ for the
colour matrices in the fundamental representation, 
to reduce the colour factor
of each Feynman diagram into combination of a linearly independent
set of `pseudo-planar'
colour factors, as illustrated in Fig.~1. 
Choose from this set all the {\it primitive colour factors}
$G_i$,
which are those that remain connected when the two quark lines are removed from
the diagram. Of those shown in Fig.~1, $G_a$ and $G_d$ are primitive,
and $G_b,G_c,G_e,G_f$ are not.
Since many Feynman diagrams can contribute to the same primitive
colour factor $G_i$,
the corresponding amplitude $d_i(s,b)$ in the impact-parameter
space will generally be given by a sum of many Feynman diagrams.

The phase shift is obtained by summing the contributions from all primitive
colour diagrams:
\be
2\delta(s,b)=\sum_id_i(s,b)G_i.\label{ps}\ee

It is clear from (\ref{impact}) that a
$\delta(s,b)$ calculate even in the lowest order will give rise to an
expression for the amplitude $A(s,\Delta)$ involving
amplitudes of all orders. In this way
a phase shift calculation of an amplitude
is similar to a renormalization-group improved calculation
of a Green's function or a physical parameter.
Moreover, it can also be shown~\cite{FL} that 
the $g^2$ and $\ln s$ dependences of $d_i(s,b)$ are of the 
form $g^{2m}D(g^2\ln s)$, if $G_i$ has $m$ gluon lines connecting
the two quarks ($m=1,2,3,2,4,4$ respectively for (a), (b), (c), (d),
(e) in Fig.~1). Let us illustrate the significance of these two remarks.

Suppose we calculate
electron-electron scattering in QED, via the exchange of $n$ photons.
We will ignore all fermion loops in the intermediate states.
There are $n!$ (crossed and uncrossed)
ladder diagrams describing this process, whose individual
powers of $\ln s$ grow with $n$. However, it is known that
when the $n!$ diagrams
are summed, {\it all} powers of $\ln s$
cancel one other, leaving behind a contribution to the total cross section
independent of energy. This can be seen from (\ref{ps}) in the following way.
In this case, 
the only primitive `colour' diagram is a single-photon exchange diagram, 
$G_a$, for all the others 
are disconnected once the two electron lines are removed
($G_d$ is not present for QED). Among the $n!$
ladder diagrams, the only one that would contribute to $G_a$ is the 
Born amplitude, with $n=1$. Hence the phase
shift is given by the Born amplitude, which is proportional to $g^2$
and contains no $\ln s$. This agrees with the general form discussed
above with $m=1$. 
Moreover, when we expand the exponential in (\ref{impact})
to order $g^{2n}$, we see that the amplitude to that order must also contain
no $\ln s$ terms, agreeing with the result obtained by direct calculations.

\begin{figure}
\vskip7.3cm
\includegraphics{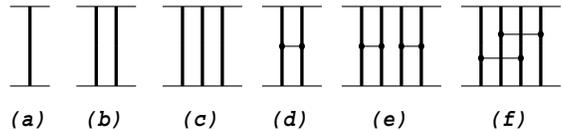}
\vspace*{-5cm}
\caption[]{Colour diagrams. The colour factors (a) and (d) are primitive,
the others are not.}
\end{figure}

For quark-quark scattering in QCD, gluon self coupling
allows an infinite number of primitive colour factors to be
formed, even when quark loops are omitted. $G_a$ is colourwise a single
gluon exchange, with amplitude of the form $d_a(s,b)\sim g^2D(g^2\ln s)$.
When all diagrams are summed in LLA, this gives rise to the familiar
Reggeized gluon amplitude with $D$ being an exponential function.
Otherwise $D$ is a polynomial if the sum is taken only to a finite
order. When the exponential in (\ref{impact}) is expanded
with $\delta(s,b)\sim d_aG_a$ being the exponent, the constant term
is cancelled and the linear term has colour-octet exchange, so it is
starting from the quadratic term that it may contribute to the
colour-singlet amplitude $A_0$ in (\ref{opt}). For that reason the
contribution from $G_a$ to $\sigma_{tot}$, is of order $g^4$, times
a function of $g^2\ln s$. In the region where $g^2$ is small and
$g^2\ln s$ is of order unity, the contribution from $G_a$ to
$\sigma_{tot}$ is then $O(g^4)$. The only other terms of the same
order came from primitive colour factors that have two vertical
gluon lines between the two quarks ($m=2$), such as $G_d$
in Fig.~1. Those diagrams are also
of $O(g^4)$, and since two vertical gluons are involved, the primitive
colour factors do contribute directly to a colour-singlet exchange.
The BFKL Pomeron amplitude is therefore the $O(g^4)$ contributions
from the $m=1$ and the $m=2$ primitive colour factors. To be sure,
other terms in the expansions of the exponential give higher order 
contributions, but these subleading-log terms are necessary to uphold
$s$-channel unitarity and the Froissart bound.

\section{Two-Loop Phase Shifts}
The quark-quark scattering amplitude in LLA, up to two loops or $O(g^6)$, can be found in the
book of Cheng and Wu~\cite{CW}.  All we need to do is to extract from it the phase shift
$\delta(s,b)$ using eq.~(\ref{ps}), and then the colour-singlet exchange amplitude $A_0$
 in (\ref{opt}). The result for the universal
energy function $f(s')$ in (\ref{f}) is given by~\cite{PH1}
\be &&f(s')=
\int d^2x\biggl[1-{2\over 3}e^{-D}
\cos\({2\delta_a\over 3}\)
-{1\over 3}e^{-2D}
\cos\({4\delta_a\over 3}\)\biggr]\nonumber\\
&&2\delta_a(s,b)=-g^2K_0(x)+(g^4N_c\ln s'/4\pi)K_0^2(x)\nn
&&\hskip3cm -(g^6N_c^2\ln^2s'/32\pi^2)V(x),\nn
&&D(x)=-i\delta_d(x)=
(g^6\ln s'/4\pi)[K_0^3(x)-\h V(x)],\label{f2}
\ee
where $K_0(x)$ is the Bessel $K$-function, and $V(x)=-(1/2\pi)^4\nabla_x^2
\int d^2x'K_0^2(|\vec x-\vec x'|)K_0^2(|\vec x'|)$.

\section{Comparison with the HERA Data}

Equation (\ref{f2}) is computed for quark-quark scattering, but if
the energy function $f(s')$ is indeed universal, as experimental
data indicate,
(see the discussion leading to (\ref{f})), 
then we can use it to compare with experimental results.
Even so, the perturbative formula in (\ref{f2}) is expected to be
quantitatively accurate
only when both $g^2$ and $g^2\ln s'$ are small.
The former is so when the virtuality $Q^2$ is large, but the latter 
is not satisfied for the present data. Higher-order calculations
are needed and they are underway~\cite{PH2}, but for now, to get
a qualitative idea, let us compare (\ref{f2}) with data anyhow. 
Choosing a linear dependence for the 
ennergy scale parameter, $\Lambda(Q)=\Lambda_0+cQ$, with $\Lambda_0=
0.2$ GeV and $c=4$, the comparison is shown in Fig.~1. We see that
inspite of the small $Q$ available in the data, the agreement 
of theory and data seems to be very good.

\begin{figure}[ht]
\includegraphics{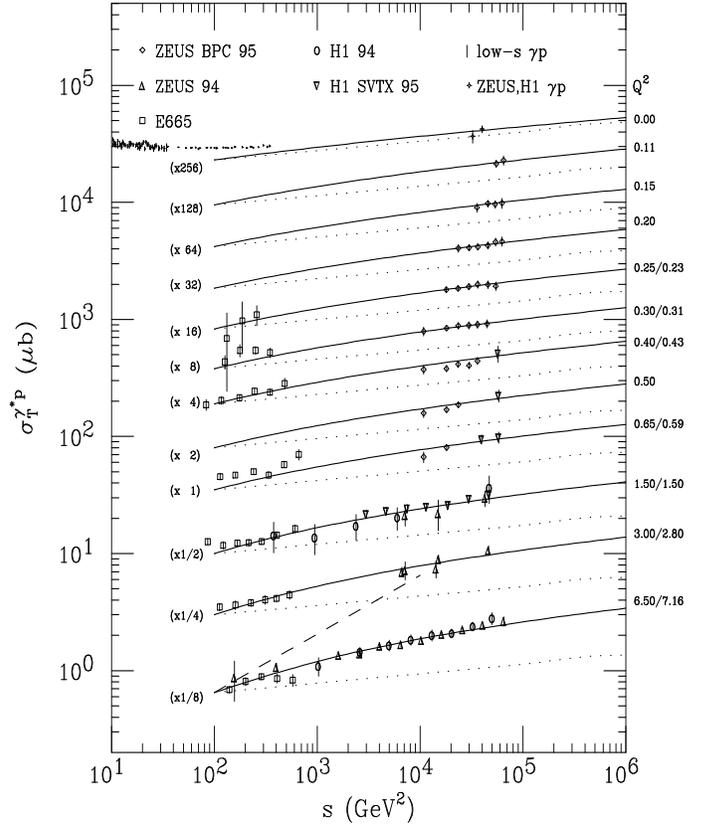}
\vskip11.3cm
\caption[]{
$\gamma^*$-proton total cross-sections as a function of $s$.
Data are taken from Ref.~\protect\cite{ZEUS} where  references
to the original experiments can be
found. $Q^2$ of the photon, in (GeV)$^2$, are given on the right.
Where two numbers are listed, the second one refers to the low-energy
data of E665. The solid curve is the prediction of the 
two-loop phase shift calculation, eq.~(\ref{f2}). The dotted lines
are $s^{0.08}$, and the dash line is $s^{0.05}$, both there for
comparisons.}
\end{figure}

 This agreement
can probably be attributed to two things. First, the mechanism explained
in section 1.2 is so general that any reasonable function $f(s')$ satisfying
the Froissart bound is expected to yield a reasonable result, if a
suitable energy-scale parameter $\Lambda(Q)$ is chosen.
Moreover, the agreement is reminiscent of the success in the
early days of QCD, before
higher-loop formulas became available. Drell-Yan pair mass spectrum
and other quantities were correctly predicted, though normalizations 
were off by the famous `K factor'. When higher-order corrections are
taken into account, there will be a correction to the universal energy
function $f(s')$ with a corresponding change in the energy-scale
parameter $\Lambda(Q)$ to fit the data.
This is not unlike $\Lambda_{QCD}$,
which takes on different values when deduced from experimentally measured
$\alpha_s(Q)$, depending on whether it appears in a one-loop or a two-loop
formula.

\section{Conclusion}
Experimentally
$\gamma^*p$ total cross section grows with energy like $s^{a(Q)}$,
with an effective power $a(Q)$ increasing with the virtuality $Q^2$
of the photon $\gamma^*$. This behaviour can be explained by universality
and the Froissart bound. To compute the universal energy function $f(s')$
governing the energy dependence of all total cross sections, we resort to
the impact-parameter formalism which preserves $s$-channel unitarity and
the Froissart bound. We have discussed a recipe in perturbative QCD
to calculate the phase shift
$\delta(s,b)$ needed in that formalism. We have also discussed
the advantage of calculating the elastic amplitude using phase shift,
rather than doing it directly from summing all Feynman diagrams.
Formulas are given for quark-quark phase shifts to two loops,
and for the corresponding energy function $f(s')$. When they are used
to compare with the HERA experimental data, 
good agreements are obtained, the
reasons for which are also discussed.

\end{document}